\documentclass[5p,times]{elsarticle}
\usepackage{epsfig}
\usepackage{nonfloat}

\usepackage{graphicx}
\usepackage{amssymb}
\usepackage[figuresright]{rotating}
\usepackage{amssymb}
\usepackage{hyperref}

\newcommand\myfigure[1]{%
\medskip\noindent\begin{minipage}{\columnwidth}
\centering%
#1%
\end{minipage}\medskip}

\begin{document}
\begin{frontmatter}
\title{Collectivity in high-multiplicity events of proton-proton collisions in the framework of String Percolation}
\author{\it{I. Bautista$^1$\thanks{irais.bautista@fcfm.buap.mx}, Arturo Fernandez$^1$, Premomoy Ghosh$^2$}}

\address{$^1$Facultad de Ciencias F\'isico Matem\'aticas, Benem\'erita Universidad Aut\'onoma de Puebla,1152, M\'exico}  
\address{$^2$Variable Energy Cyclotron Centre, 1/AF Bidhannagar, Kolkata 700 064, India}  
\date{}
\begin{abstract}
We analyze high multiplicity proton-proton ($pp$) collision data at the energies $\sqrt{s}=900$ GeV, $2.76$ TeV and $7$ TeV in the framework of the String Percolation Model (SPM) in terms of the ratio of shear viscosity and entropy density ($\eta/s$) showing that the model allows the formation of strongly interacting collective medium in the high energy and high multiplicity events in p-p collisions.

\end{abstract}

\end{frontmatter}
\section{Introduction}
In the search for signals for the Quark-Gluon Plasma (QGP) in heavy-ion experiments, one estimates possible effects of incoherent 
superposition of binary nucleon-nucleon collisions \cite{ref09}, that serves the baseline for extracting the signals, while the $pp$ collision data, at respective 
centre-of-mass energy, provide the required input on cross-section of nucleon-nucleon collisions in modeling the multiparticle production by binary collisions 
among the participating nucleons of the colliding nuclei. In this scenario of the QGP study, an unexpected feature, namely the "ridge" in the long range near side
angular correlation, in distinct class of "high multiplicity" events of proton-proton collisions \cite {ref10} at $\sqrt {s} = $ 7 TeV at LHC has triggered revival of an old 
school of thought  \cite {ref11, ref12, ref13, ref14} of the possibility of formation of a collective medium in $pp$ collisions also. Several of subsequent theoretical 
and phenomenological studies \cite {ref14a, ref15, ref16, ref16,ref17}, in different approaches, endorse the possibility, indicating the need for further investigations 
in understanding the high-multiplicity $pp$ events vis-a-vis the QGP. In this article, we address the issue of collectivity in high multiplicity $pp$ 
events in the framework of the String Percolation Model, that has successfully explained the collectivity and the change of phase in nucleus-nucleus collisions 
\cite{ref19, ref20, ref21}. Most importantly, in the context of the present work, the SPM also describes the centre-of-mass energy dependence of mid-rapidity 
multiplicity \cite {ref21a} and the pseudorapidity distributions \cite {ref21b} of produced charged particles in $pp$ collisions, for the entire range of energy, available so far.  

 \section{Multiparticle Production in String Percolation Model}
In the String Percolation Model (SPM), the sources of multiparticle productions are the color strings between the colliding partons. The stretched strings between the receding partons decay into new pair of partons and so new strings are formed. Subsequently, particles are produced from interaction of partons
by the Schwinger Mechanism \cite{ref21c}. The higher the energy of the collision, the more dominant becomes the role of the sea quarks and gluons, resulting 
in availability of a large number of color strings. So, with the increasing energy of collision and / or the size of the colliding system, the density of the string 
increases and they start to overlap forming clusters. The overlapped strings start interacting. At a certain critical density, the strings start percolating through one 
another, forming a macroscopic cluster of strings - a geometrical phase transition takes place. The cluster of percolated color strings is considered to be equivalent 
to the de-confined partonic  state of matter. In fact, there has been considerable progress in the SPM in establishing connection  \cite{ref20, ref21, ref22, ref23, ref24} 
between the percolation phase transition of color strings and the QCD phase transition in heavy-ion collisions. As already discussed, the model has been extended 
in describing features of multiparticle production in $pp$ collisions also.

One of the relevant parameters in SPM is the transverse impact parameter density of strings, $\zeta^{t}$. For pp collisions, \cite{ref25, ref26} one can write
\begin{equation}
\zeta^{t}\equiv(\frac{r_{0}}{R_{p}})^{2}\bar{N}^{s}
\end{equation}
where $r_{0}$ is $\simeq 0.25$ fm the single string transverse size, $R_{p}\simeq 1$ fm is the proton transverse size and $\bar{N}^{s}$ is the average
number of single strings.  For values of $\zeta^{t}$ below the critical value for the 2-dimensional percolation, $\zeta^{t}_{c}$ ($\simeq 1.2- 1.5$, depending on the profile function homogeneous or Wood Saxon type), the strings do not interact and the collective effect is absent.
For values $\zeta^{t} \ge \zeta^{t c}$, one observes the formation of long strings due to fusion and stretching between the colliding nucleons. 
To search for de-confinement and collectivity, we consider $\zeta^{t} \geq \zeta^{t}_{c}$. 

The other principal parameter in SPM is the Color Suppression Factor \cite{ref27, ref28}, F($\zeta^{t}$), which is related to the particle density $dN/dy$
and the number ($\overline{N}^{s}$) of strings as:
\begin{equation}
\frac{dN}{dy}=\kappa F(\zeta^{t})\bar{N}^{s}
\end{equation}
where $\kappa$ is a normalization factor $\sim .63$ \cite{ref29} and
\begin{equation}
F(\zeta^{t})\equiv \sqrt{\frac{1-e^{-\zeta^{t}}}{\zeta^{t}}}
\end{equation}

For pp collisions one can approximately write 
$N_{p}^{s}=2+4(\frac{r_0}{R_{p}})^{2} e^{2\lambda Y}$
with $Y$ the beam rapidity and $\lambda$ a constant parameter $\simeq.201$ \cite{ref29}.

In the above equation one may notice that the factor $F(\zeta^{t})$ slows down the rate of increase in particle density with energy and with the number of strings, 
as the strength of the color field inside a cluster of n strings is $\sqrt{n}$ and the strength of a single string is due to the random direction of the individual color 
field in color space. 

The Schwinger model with percolation \cite{ref29a}
\begin{equation}
\frac{dN}{dp_{T}}\sim e^{- p_{T} \sqrt{ \frac{2 F(\zeta^{t})}{\langle p_{T} \rangle _{1}}}},
\end{equation}

 \section{ Analysis of high multiplicity $pp$ data in SPM}
 
 The CMS experiment at the LHC has measured $p_{T}$-spectra from \cite{ref33, ref34} pions ($\pi^\pm$), kaons ($K^\pm$) and protons  ($p$ and $\bar p$) from $pp$ 
 collisions at $\sqrt {s}$ = 0.9, 2.76 and 7 TeV in the rapidity range $|y|<1$ for different classes of events depending on mean number of charged particles, 
 $\langle N_{ch} \rangle$ in the pseudo-rapidity interval, $|\eta| <2.4$. The measured $p_{T}$ - ranges are  (0.1 to 1.2) GeV/c for $\pi^\pm$, (0.2 to 1.050) GeV/c 
 for $K^\pm$ and (0.35 - 1.7) GeV/c for $p$ and $\bar p$. By measuring the $d\langle N_{ch} \rangle /d\eta$ - dependent identified particle spectra, the CMS 
 experiment provides the unique opportunity for somewhat  "centrality" dependent study in $pp$ collisions. We fit the transverse momentum distributions for identified particles from different classes of high multiplicity events to obtain the string density, to evaluate the corresponding $\eta/s$, (measure of fluidity) to check the collective 
 nature of the medium, if formed.   
 
The distribution of transverse momentum for any type of collision can be expressed as \cite{ref34a}
\begin{equation}
\frac{d^{2}N}{dp_{T}^{2}}=\frac{a'}{[p_{0}\sqrt{\frac{F(\zeta_{pp})}{F(\zeta_{HM})}+p_{T}}]^{\alpha}}
\end{equation}
so that the transverse momentum distribution can be write as
\begin{equation}
\frac{1}{N} \frac{d^{2}N}{d\eta dp_{T}}=\frac{a
(p_{0}\frac{F(\zeta_{pp})}{F(\zeta_{HM})})^{\alpha-2}
}{[p_{0}\sqrt{\frac{F(\zeta_{pp})}{F(\zeta_{HM})}+p_{T}}]^{\alpha-1}}
\end{equation}

In the case of pp collision with no high multiplicity data (min bias) equation (5) becomes 
\begin{equation}
\frac{d^{2}N}{dp_{T}}=\frac{a}{[p_{0}+p_{T}]^{\alpha}}
\end{equation}
and so equation (6) transform into 
\begin{equation}
\frac{1}{N} \frac{d^{2}N}{d\eta dp_{T}}=\frac{a p_{0}^{\alpha-2}
}{[p_{0}+p_{T}]^{\alpha-1}}
\end{equation}

Using the high multiplicity data of pions from reference \cite{ref33} we had perform a fit over the in the different collision energies  transverse momentum distribution of produced particles with equation (8), where $a$, $\alpha$ and $p_{0}$ are energy dependent parameters of the experimental observed spectra in p-p collisions at a given energy, $\zeta_{pp}$ and $\zeta_{HM}$ are the percolation parameters for the p-p collisions and the high multiplicity p-p collisions respectively. All these parameters are energy dependent but below 200 GeV $\zeta_{pp}$  is small so that $F(\zeta^{t})\simeq 1$.

Table 1. contains the obtained corresponding parameters from the fit of the equation(6) with data \cite{ref34}. 
\begin{table}[h]
\centering
\begin{tabular}{|c|c|c|c|}
\hline
$\sqrt{s}$  (TeV) & a & $p_{0}$ & $\alpha$ \\
\hline
.9 & 23.29 $\pm$ 4.48 & 1.82 $\pm$ .54 & 9.40 $\pm$ 1.79 \\
\hline
2.76 & 22.48 $\pm$ 4.20 & 1.54 $\pm$ .46 & 7.94 $\pm$ 1.41\\
\hline
7 & 33.11 $\pm$ 9.31 & 2.31 $\pm$ .87 & 9.78 $\pm$ 2.53 \\
\hline
\end{tabular}
\caption {Parameters of the transverse momentum distribution (6) in pp collisions.}
\end{table}

Now by using the obtained parameters we can perform the fit over the high multiplicity data by using equation (6) and obtain the corresponding $\zeta_{HM}$ (Figure 1). We used the data of pions where we have restrict the fit to the $p_{T} > .4$ to avoid the effect of resonance decays. 
For the values of $dN/d\eta$, here we have taken into the account the kinematics cuts, by scaling the measured $\langle N_{track} \rangle$ with the corresponding factor of ($1/4.8$) corresponding to the $|\eta|<2.4$ range and the factor ($1.6$ ) as in \cite{ref35a} corresponding to the $p_{T}> 0.4$ GeV$/c$ cut range. For each value of $dN/d\eta$ we have calculated the corresponding $\zeta^{t}$ as shown in the following table.
\begin{table}[h]
\centering
\begin{tabular}{|c|c|c|}
\hline
$\langle N_{track} \rangle$ & $dN/d\eta$ & $\zeta^{t}$ \\
\hline
40 & 13.33 & .77 $\pm$ .13 \\
\hline
52 &17.33 & 1.42 $\pm$ .15\\
\hline
63 & 21.0 & 1.98  $\pm$ .18\\
\hline
75 & 25  & 2.53 $\pm$ .21\\
\hline
86 & 28.67 & 3.02 $\pm$ .24\\
\hline
98 & 32.67 &  3.42  $\pm$ .26\\
\hline
109 & 36.33  & 3.89 $\pm$ .30\\
\hline
120 & 40.  &  4.40 $\pm$ .36\\
\hline
131 & 43.67 & 4.98 $\pm$  .40\\
\hline

\end{tabular}
\caption { $ \langle N_{track} \rangle$, $dN/d \eta$ and $\zeta ^{t}$ in pp collisions at 7 TeV high multiplicity clases.}
\end{table}

\myfigure{\includegraphics[width=1\columnwidth]{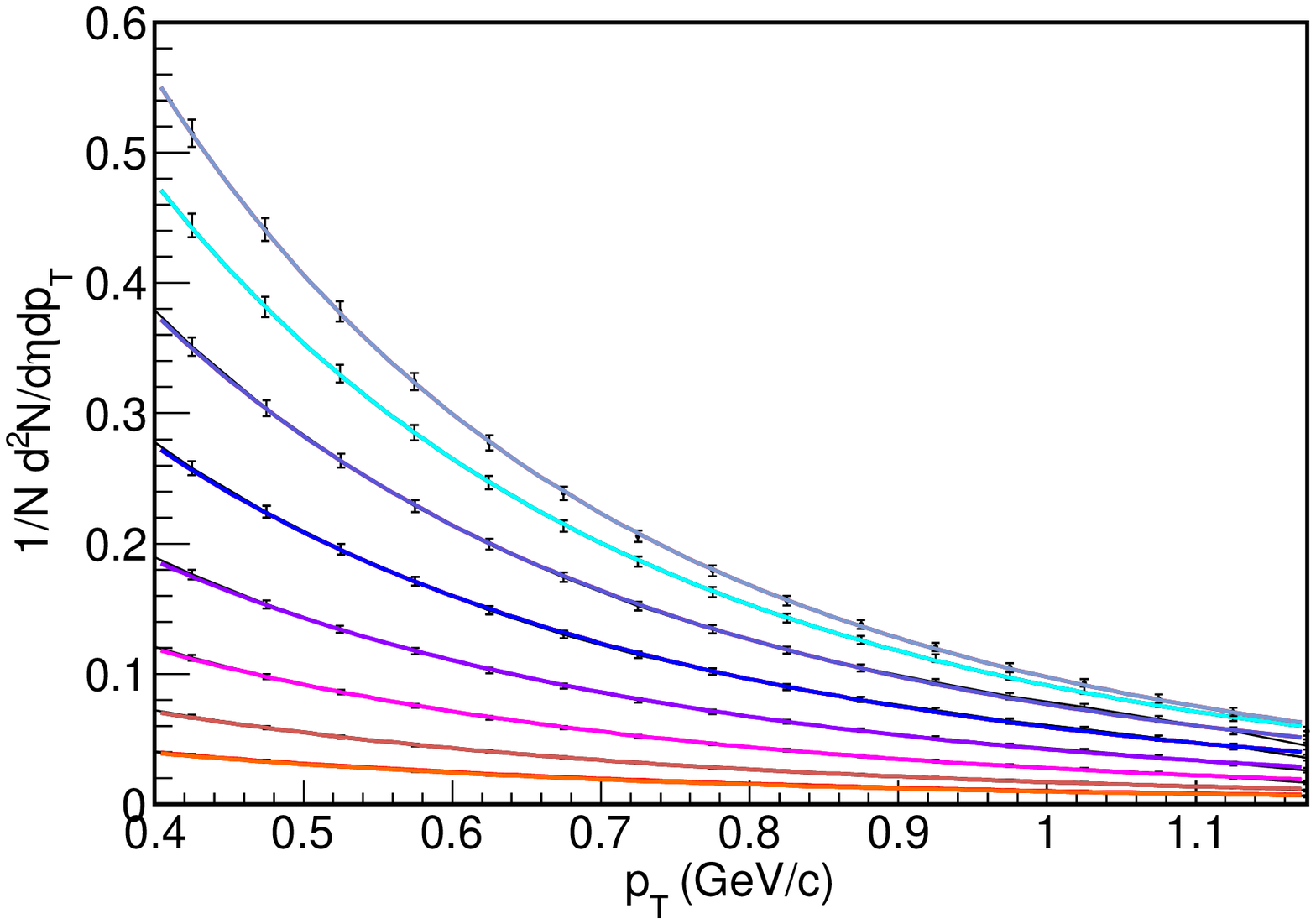}
\figcaption{Fits to the transverse momentum distribution for 7 TeV $p-p$ collisions for different multiplicity classes from $N_{track}=40$ grey line to $N_{track}=131$ orange line. Data taken from reference \cite{ref34}.}}

Figures 2 and 3 shown the dependence of $\zeta^{t}$ and $F(\zeta^{t})$ with the corresponding $dn/d \eta$ respectively at different energies.

\myfigure{\includegraphics[width=1\columnwidth]{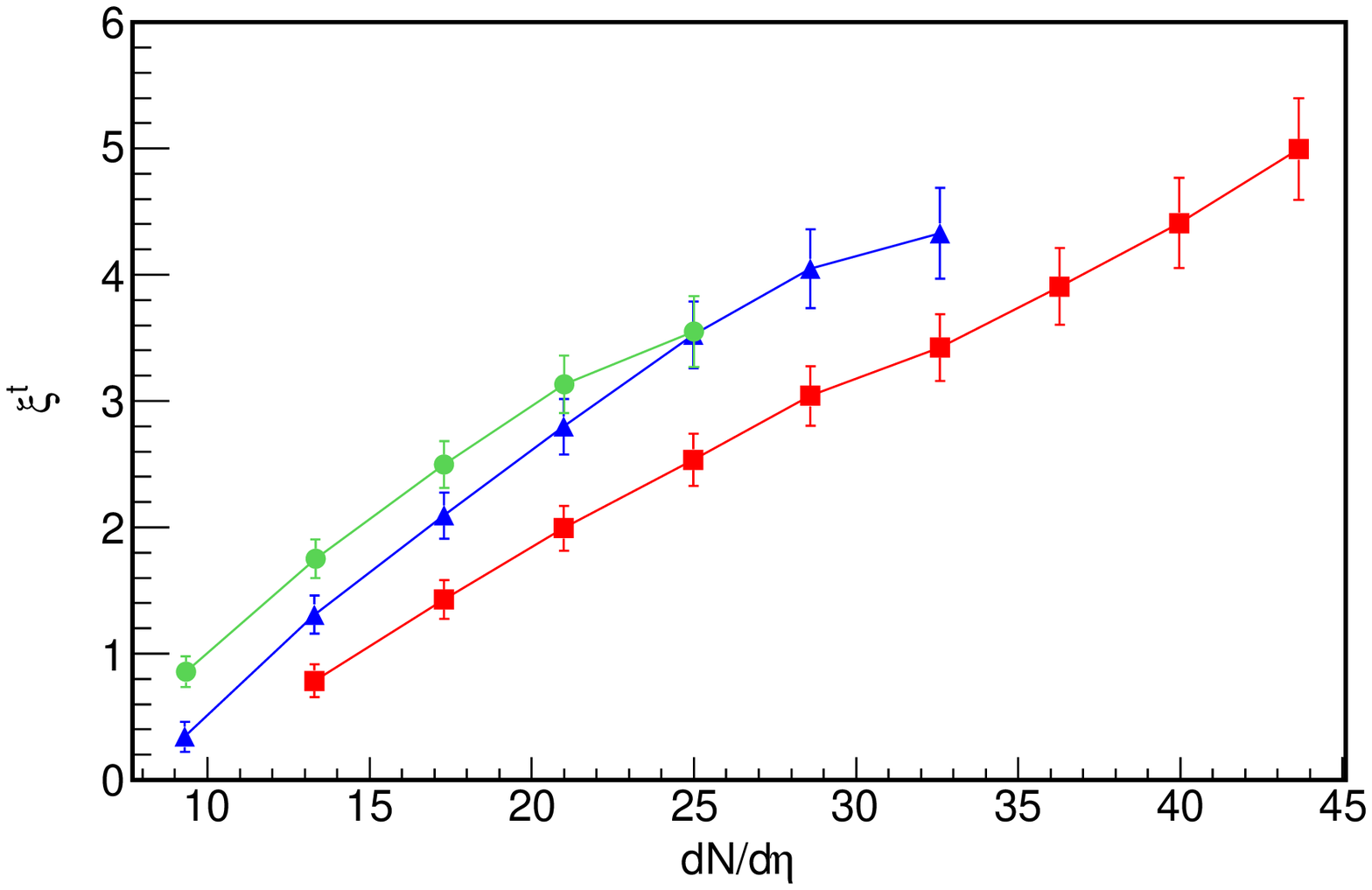}%
\figcaption{$\zeta^{t}$ for high multiplicity clases at different energies.}}
\myfigure{\includegraphics[width=1\columnwidth]{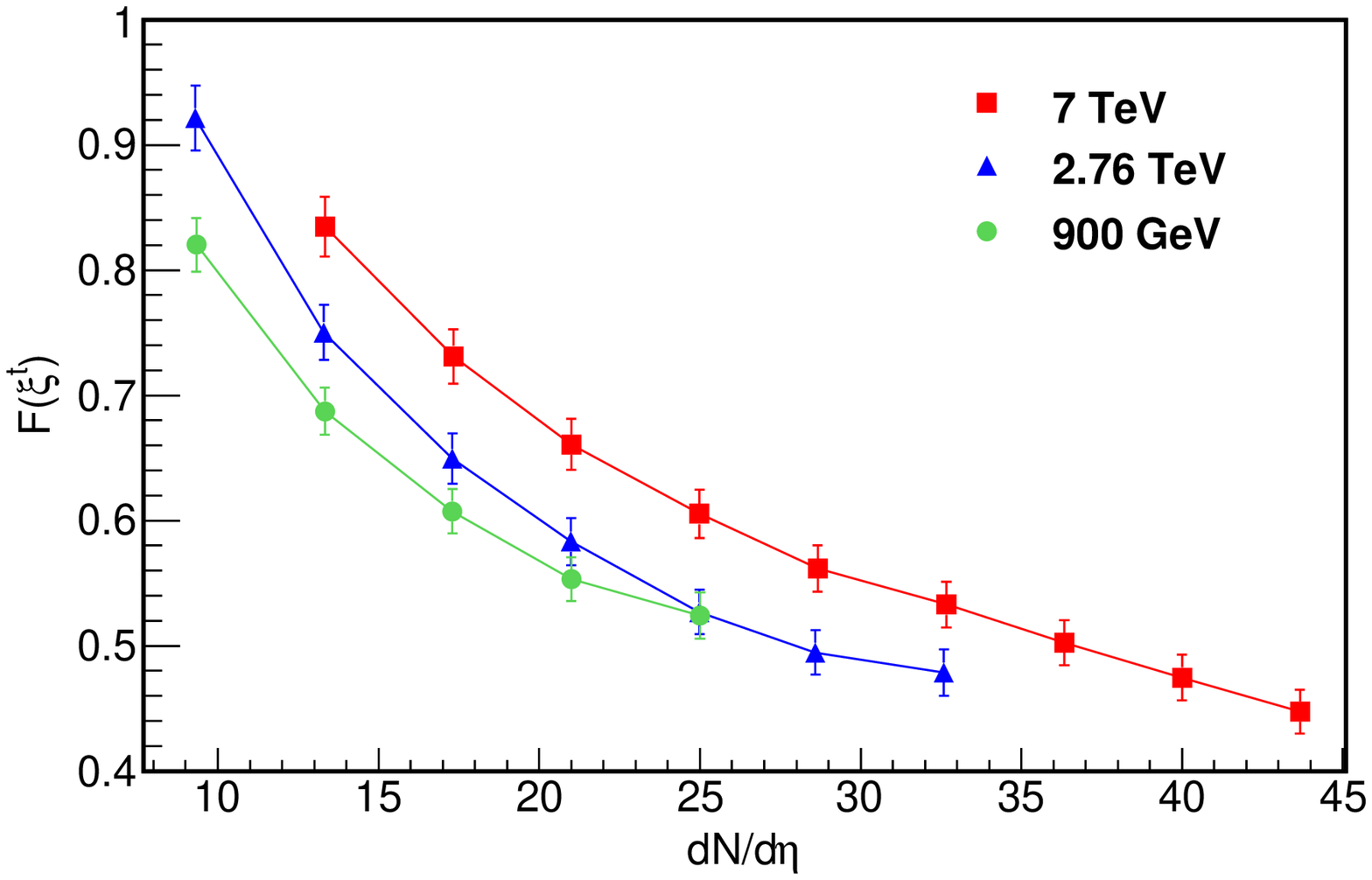}%
\figcaption{Color reduction factor at high multiplicities for different energies}}

 \section {Shear Viscosity over Entropy Density ($\eta/ s$)}

The equation (4) is the Schwinger mechanism for massless particles which can be related to the average value of the string tension 
\begin{equation}
\langle x^{2} \rangle =\pi \langle p_{T}^{2}  \rangle_{1} /F(\zeta)
\end{equation}
this value fluctuates around its mean value because the chromoelectric field is not constant, the fluctuations of the chromo electric field strength lead to a Gaussian distribution of the string tension that transform it into a thermal distribution, where the temperature is been given by the relation \cite{ref19}
\begin{equation}
T(\zeta^{t})=\sqrt{\frac{\langle p_{T}^{2} \rangle _{1}}{2F(\zeta^{t})}}
\end{equation}

We consider that the experimental, determined chemical freeze out temperature is a good mesure of the phase transition temperature $T_{c}$. 

The single string average transverse momentum $\langle p_{t} \rangle _{1}$ is calculated at 
$\zeta_{c}=1.2$ and $\zeta_{c}=1.5$ with the universal chemical freeze out temperature of $167.7\pm 2.6$ MeV and $154\pm 9$ MeV both values corresponding to the old and new LQCD results from the HotLQCD collaboration \cite{ref22a}. The values are given in table 3 which are close to $\sim 200$ MeV.

\begin{table}[h]
\centering
\begin{tabular}{|c|c|c|}
\hline
 $T_{c}$  &  $\zeta^{t}_{c}$ & $p_{T 1}$ \\
 \hline
154 $\pm 9$ &  $1.2$  & $190.25$ $\pm 11.12$  \\
\hline
154 $\pm 9$ &  $1.5$  & $184.76$ $\pm 7.80$  \\
 \hline
167.7 $\pm 2.3$ &  $1.2$  & $207.18$ $\pm 3.21$  \\
 \hline
167.7 $\pm 2.3$ &  $1.5$  & $201.19$ $\pm 3.12$  \\
 \hline
\end{tabular}
\caption {Temperature, and string density and $p_{T 1}$ in pp collisions }
\end{table}

Above the $\zeta_{c}$ the size and density spanning cluster increases. We compare the obtained temperature $T_{i}$ at the measured value of $\zeta=2.88$ before the expansion of the QGP with the measured $T_{i}=221 \pm 19_{stat} \pm 19_{sys}$ MeV from the enhanced direct photon experiment measured by PHENIX \cite{ref22b}.

We observed that all the values used in table 3 are 
consistent with the previously used value of $\sim 200$ MeV \cite{ref22d} in the calculation of percolation transition of temperature with the exception of the one obtained at $T_{c}=154$ with $\zeta_{c}=1.5$.

To calculate the effective temperature for each multiplicity for the critical density $\zeta_{c}=1.2$ and a critical temperature $T_{c}=154 \pm 9$ MeV, with the corresponding $\langle p_{T} \rangle_{1}$ we used equation (10) to get the value of $T$ from a given string density $\zeta^{t}$ which correspond to a mean number of produced strings in a given collision energy. 

\begin{table}[h]
\centering
\begin{tabular}{|c|c|}
\hline
$dn/d\eta$ & $T$ \\
\hline
   5.88 &  145.86 $\pm$  8.52\\
\hline
   13.33   &     147.41  $\pm$     10.71 \\
   \hline
   17.33    &    157.51 $\pm$    11.54 \\
   \hline
      21.00     &   165.71 $\pm$   12.22 \\
  \hline
   25.00      &  173.37 $\pm$   12.88 \\
  \hline
   28.67     &  179.75 $\pm$       13.45 \\
  \hline
   32.67     &  184.55   $\pm$       13.92 \\
  \hline
   36.33     &   190.09     $\pm$   14.50 \\
  \hline
   40.00   &  195.61  $\pm$      15.17 \\
  \hline
   43.67  &  201.46    $\pm$     15.69 \\
\hline
\end{tabular}
\caption { Obtained $T$ for different multiplicity classes at 7 TeV. }
\end{table}

Table 4 and Figure 4 show the growing of $T$ with the increase of the multiplicity $dn/d\eta$.

\myfigure{\includegraphics[width=1\columnwidth]{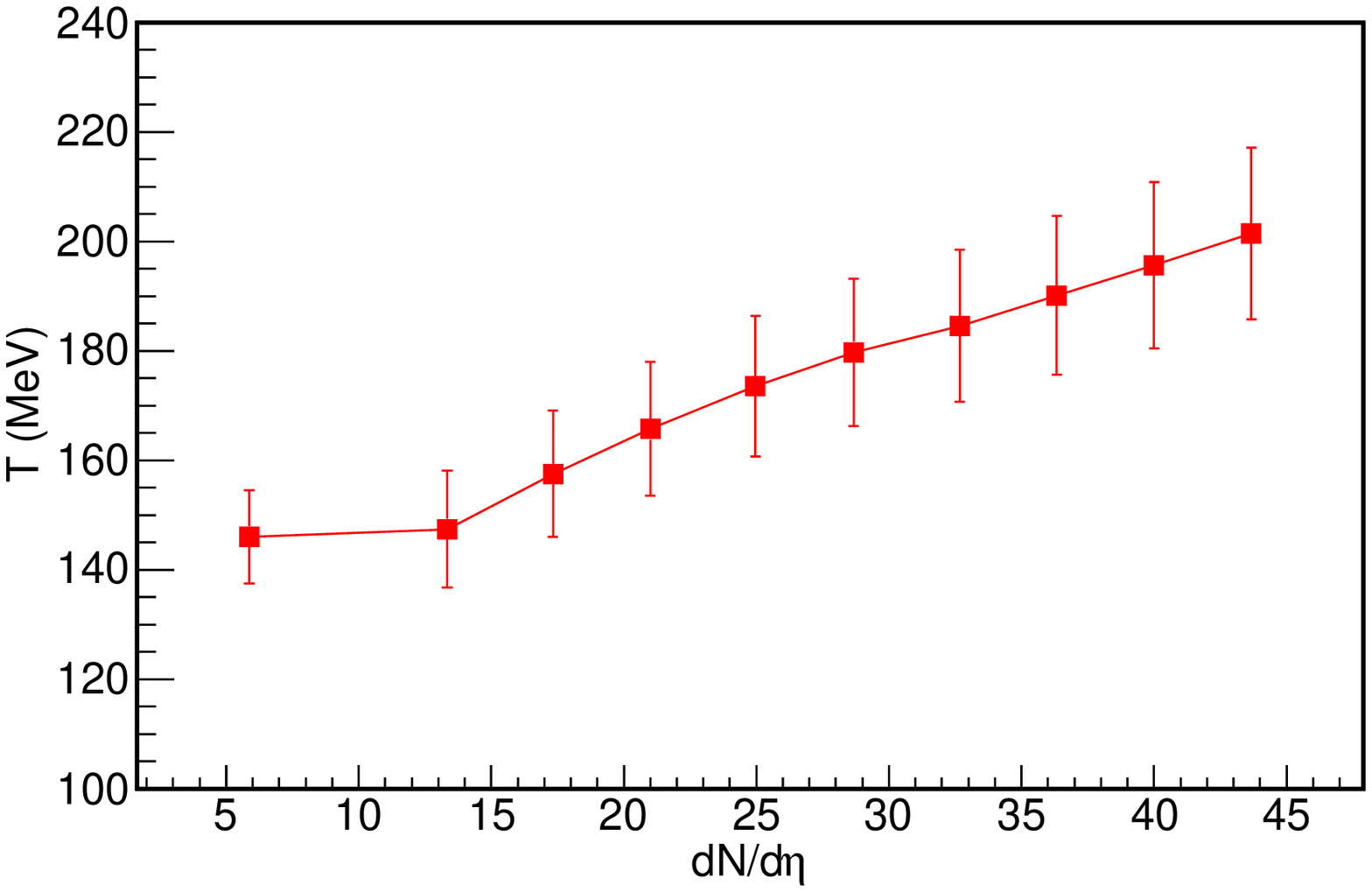}%
\figcaption{Efective temperature vs  $dN/d\eta$ for high multiplicity classes at $\sqrt{s}=7$ TeV}}

In terms of the effective temperature one can study some useful quantities as the ratio of the shear viscosity over entropy density which is given in the relativistic kinetic theory as \cite{ref22e},\cite{ref22f}
\begin{equation}
\eta/s \simeq \frac{T\lambda_{fp}}{5}
\end{equation}
where $\lambda_{mfp}$ is the mean free path $\sim\frac{1}{n\sigma_{tr}}$, n is the number of density of the effective number of sources per unit volume and $\sigma_{tr}$ is the transport cross section.
\begin{equation}
n=\frac{N_{sources}}{S_{N}L}
\end{equation}
It is considered that 
\begin{equation}
\frac{N_{sources}}{S_{N}L} \sigma_{tr} = (1-e^{-\zeta^{t}})/L
\end{equation}
considering $L=1 fm$ the longitudinal extension of the source one can give the relation $\eta/s$ in terms of $\zeta^{t}$ as in reference \cite{ref22f}

\begin{equation}
\frac{\eta}{s}=\frac{TL}{5(1-e^{-\zeta^{t}})}
\end{equation}

In figure 5 we show the dependence of $\frac{\eta}{s}$ from (12) on the ratio $T/T_{c}$. The results changes very little for the 3 different scenarios of critical values given in Table 3. 

\myfigure{\includegraphics[width=1\columnwidth]{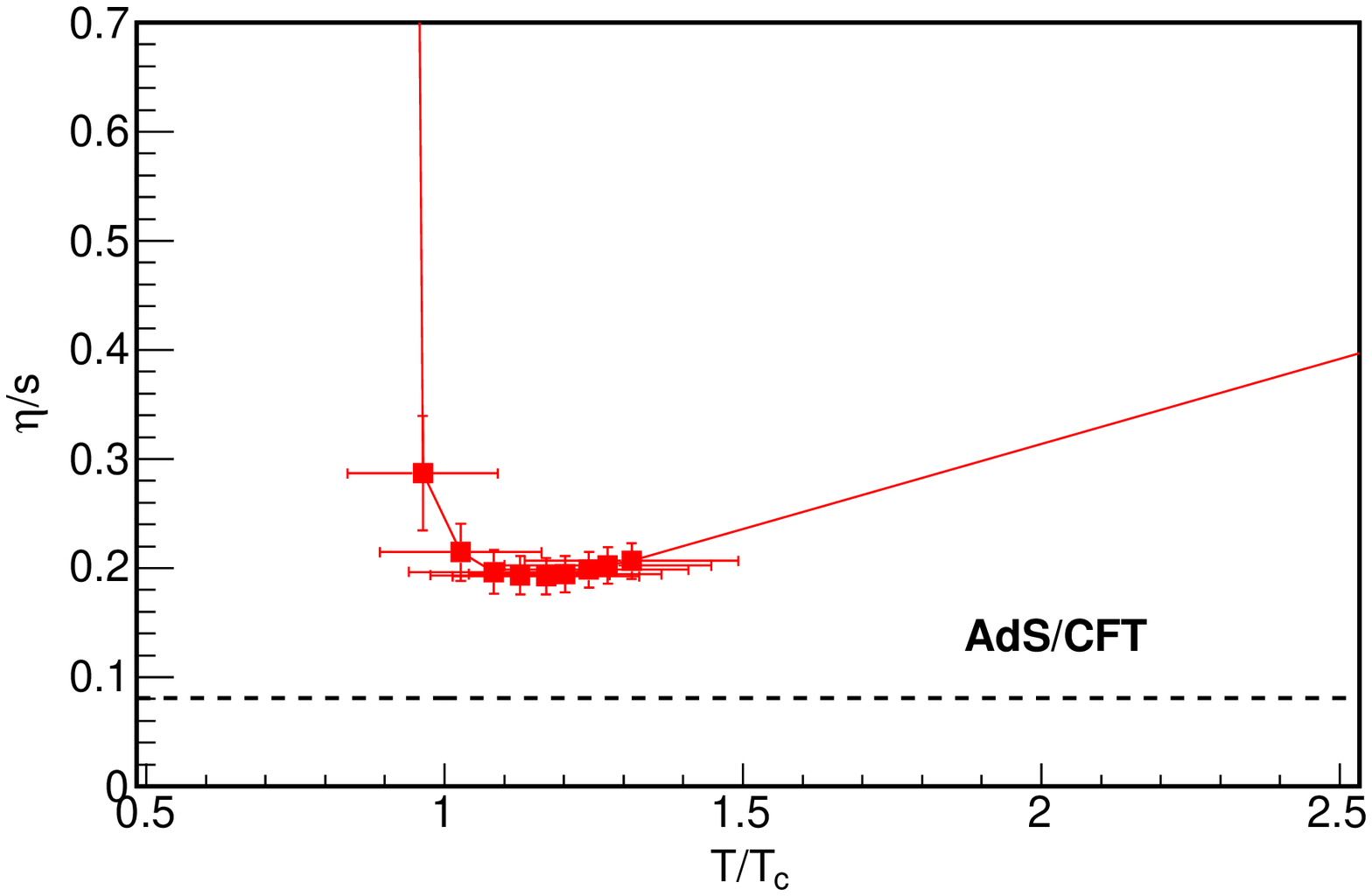}%
\figcaption{Shear viscosity over entropy ratio for 7 TeV high multiplicity classes corresponding to $N_{track}=40$ up to $N_{track}=131$, with the $T_{c}=154\pm9$. AdS/CFT result is take as $1/(4\pi)$.}}
Table 5 shows the values for the ratio $\frac{\eta}{s}$, as obtained from equation (12) for the different event clases.

 \begin{table}[h]
\centering
\begin{tabular}{|c|c|}
\hline
 $T/T_{c}$ & $\eta/s$ \\
\hline
   0.953 $\pm$  0.111 & 1.489 $\pm$  .087 \\
   \hline
  0.964  $\pm$     0.126    &  0.285   $\pm$     .052 \\
  \hline
   1.031 $\pm$     0.135    &   0.213   $\pm$     .025 \\
   \hline
   1.084 $\pm$     0.143     &   0.196     $\pm$      .020 \\
   \hline
   1.134   $\pm$      0.150  &     0.192     $\pm$    .017 \\
   \hline
   1.176 $\pm$     0.156   &    0.192   $\pm$      .017 \\
   \hline
   1.208 $\pm$       0.161   &    0.194   $\pm$    .016 \\
   \hline
   1.244  $\pm$     0.167  & 0.197    $\pm$    .016 \\
   \hline
   1.280  $\pm$      0.173   &   0.201  $\pm$      .016\\
   \hline
   1.319   $\pm$      0.179  &    0.206    $\pm$    .166 \\
   \hline
\end{tabular}
\caption {Shear viscosity over entropy ratio for 7 TeV high multiplicity classes, with the $T_{c}=154\pm9$. The fist row correspond to the values for the min bias $dn/d\eta$ in order to compare as a reference. }
\end{table}

\section{Conclusions}
We have seen in this work that the measured high multiplicity events in $p-p$ collisions at the energies $900$ GeV, $2.76$ TeV and $7$ TeV are able to reach the geometrical phase transition in the framework of string percolation model that marks the critical string density at which collective effects appears as a result of the increasing density due to a kind of centrality at these events. 

\section{Acknowledgments}
I. Bautista was supported by CONACYT grant, and thank C. Pajares, B. K. Srivastava for communication.

\end{document}